\documentclass[conference]{IEEEtran} %
\IEEEoverridecommandlockouts
\usepackage[utf8]{inputenc}

\title{ Dynamic Trajectory and Offloading Control of UAV-enabled MEC under User Mobility}

\author{\IEEEauthorblockN{Zheyuan Yang\IEEEauthorrefmark{1}, Suzhi Bi\IEEEauthorrefmark{2}, and Ying-Jun Angela Zhang\IEEEauthorrefmark{1}}   \IEEEauthorblockA{\IEEEauthorrefmark{1}Department of Information Engineering, The Chinese University of Hong Kong, Shatin, N.T., Hong Kong SAR}\IEEEauthorblockA{\IEEEauthorrefmark{2}College of Electronics and Information Engineering, Shenzhen University, Shenzhen, Guangdong, China 518060
    \\ E-mail: \{yz019, yjzhang\}@ie.cuhk.edu.hk, bsz@szu.edu.cn} }
\usepackage{cite}
\usepackage{graphicx}
\usepackage{nicefrac}
\usepackage{amsmath}
\usepackage{amssymb}
\usepackage{amsthm}
\usepackage{commath} 
\usepackage{mathtools}
\usepackage{relsize}%
\usepackage[justification=centering,font=scriptsize]{caption}
\usepackage{subfigure}
\usepackage{bm}
\usepackage{epstopdf}
\usepackage{algcompatible}
\usepackage{algorithm}

\usepackage[belowskip=-15pt,aboveskip=0pt]{caption}
 
\catcode`_=\active
\newcommand_[1]{\ensuremath{\sb{\scriptscriptstyle #1}}}

\catcode`^=\active
\newcommand^[1]{\ensuremath{\sp{\scriptscriptstyle #1}}}
\newcommand\addtag{\refstepcounter{equation}\tag{\theequation}}

\allowdisplaybreaks
\begin{document}
\maketitle

\begin{abstract}
In this paper, we consider a UAV-enabled MEC platform that serves multiple mobile ground users with random movements and task arrivals. We aim to minimize the average weighted energy consumption of all users subject to the average UAV energy consumption and data queue stability constraints. To control the system operation in sequential time slots, we formulate the problem as a multi-stage stochastic optimization, and propose an online algorithm that optimizes the resource allocation and the UAV trajectory in each stage. We adopt Lyapunov optimization to convert the multi-stage stochastic problem into per-slot deterministic problems with much less optimizing variables. To tackle the non-convex per-slot problem, we use the successive convex approximation (SCA) technique to jointly optimize the resource allocation and the UAV movement. Simulation results show that the proposed online algorithm can satisfy the average UAV energy and queue stability constraints, and significantly outperform the other considered benchmark methods in reducing the energy consumption of ground users. 
 \end{abstract}

\section{Introduction}
\par The proliferation of mobile devices is accelerating the development of Internet of Things (IoT) and the advent of mobile applications with new intelligent features, such as automatic navigation and augmented reality/virtual reality \cite{shi2016edge}. Such applications are often computationally demanding and sensitive to latency. However, IoT devices cannot support high-performance computations on account of limited battery and low computing capability\cite{bi2018computation}. Mobile edge computing (MEC) offers a cost-effective solution to cater for computation-intensive and latency-critical tasks, by pushing computational resources towards the network edges (e.g., base stations, access points) in proximity to end users\cite{mao2017survey}. Nevertheless, for scenarios with limited infrastructure coverage, e.g., due to severe shadowing or natural disaster-caused damage, mobile devices with poor wireless connectivity cannot enjoy the benefits of MEC\cite{mozaffari2019tutorial}. Unmanned aerial vehicle (UAV) has been envisioned as an important means to assist future wireless communications\cite{zeng2016wireless}, and it can act as aerial computing platform to directly provide flexible and resilient computation services to mobile users\cite{pham2019survey}.

In many mobile computing scenarios, computation demands arrive stochastically and the user locations change dynamically. It necessitates the design of online algorithms for UAV-enabled MEC to make real-time control decisions. Considering user mobility, the authors in \cite{liu2020path} propose a double deep Q-network (DDQN) based algorithm to maximize the long-term throughput subject to UAV energy and quality of service (QoS) constraints. However, it considers an overly simplified UAV mobility model that the UAV can only hover over one of few fixed locations. Another related work \cite{zhang2018stochastic} considers stochastic user data arrivals and minimizes the long-term average weighted sum system energy under queue stability and UAV trajectory constraints. However, \cite{zhang2018stochastic} considers static ground UEs and re-computes the whole trajectory from the initial position to destination in each time slot, which incurs significant extra computational complexity. From the above discussion, none of the work above addresses the coexistence of stochastic data arrivals, user mobility and per-slot UAV movement control in their online algorithm designs.
 
 In this work, we consider a UAV-enabled aerial MEC server providing computing services to mobile ground users with stochastic movement and data arrivals. By jointly optimizing the resource allocation and UAV trajectory, we aim to design an online algorithm to minimize the time average weighted sum energy consumption of the ground users subject to the average UAV energy consumption constraint and the data queue stability constraint. We summarize our contributions below.
 
\begin{itemize}
	\item  Under both stochastic data arrival and user mobility, we formulate a multi-stage stochastic optimization problem to jointly optimize UAV trajectory and resource allocation in sequential time slots. Without prior knowledge of the system randomness, the online algorithm design is particularly difficult to satisfy the long-term queue stability and the average UAV energy consumption constraints. 
	\item To address the coupling effect of the sequential control decisions, we apply the Lyapunov optimization to decouple the multi-stage stochastic problem into per-slot deterministic optimization problems that optimize only the resource allocation and UAV movement within each time slot. 
	\item  To tackle the non-convex per-slot optimization problem, we propose a reduced-complexity method that jointly optimizes the resource allocation and UAV trajectory using the successive convex approximation (SCA). 
\end{itemize}

Simulation results show that applying the joint method to solve the per-slot problems can satisfy the average UAV energy and queue stability constraints, and the proposed method significantly outperforms the other considered benchmark methods in reducing the energy consumption of ground users.

\section{System Model and Problem formulation}
\begin{figure}
    \centering
    \includegraphics[width=0.35\textwidth]{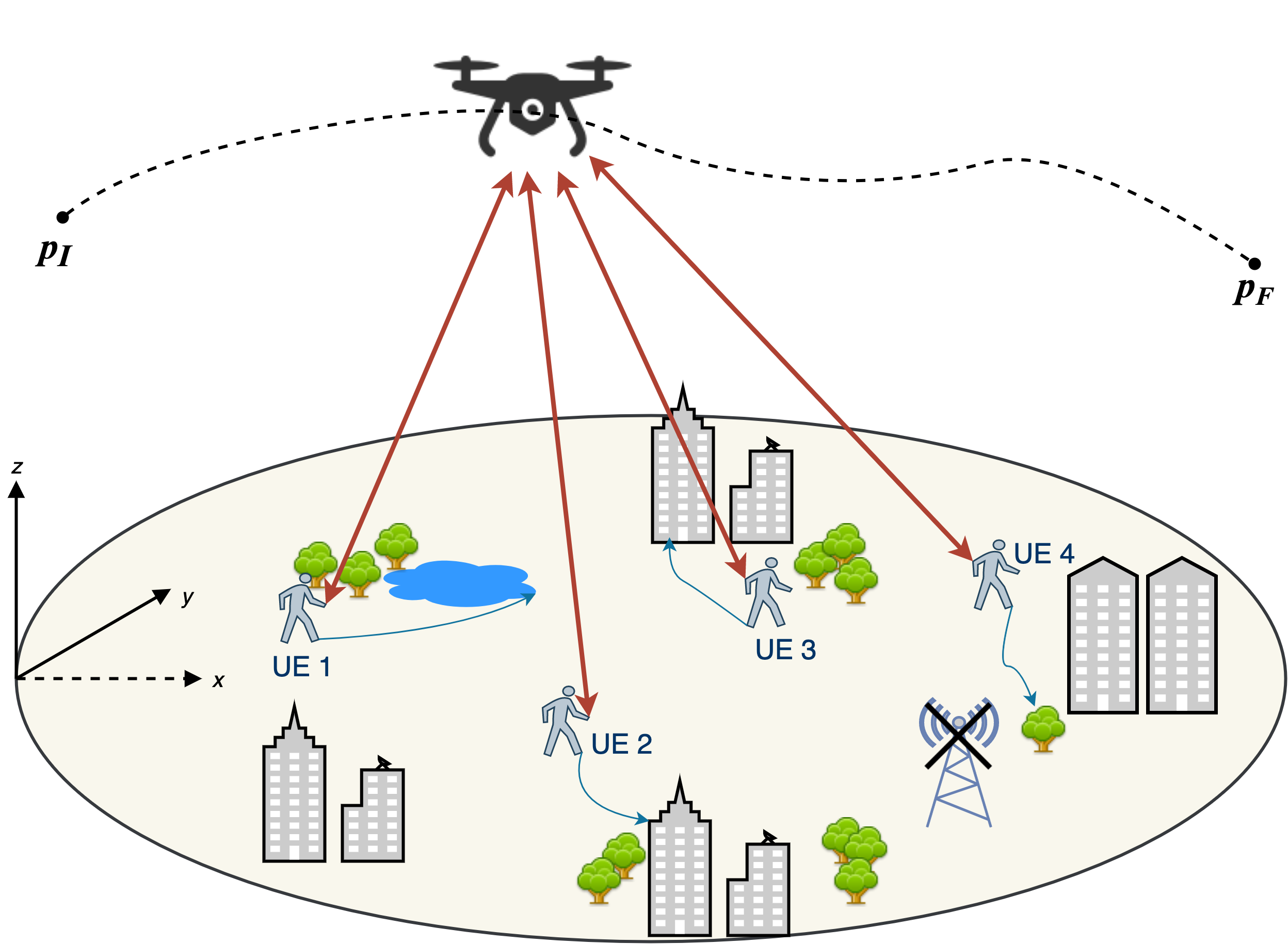}
    \caption{UAV-enabled MEC system model}
    \label{fig:sys}
\end{figure}

 As shown in Fig. \ref{fig:sys}, we consider a UAV-enabled MEC system consisting of a UAV-mounted cloudlet and a set $\mathcal{K} = \{1,2,...,K\}$ of $K$ ground user equipments (UEs). The UAV relies on limited on-board battery energy to fly above the ground users and provide edge computing service in duration $D$, which is equally discretized into a set $\mathcal{N} =\{1,...,N\}$ of $N$ time slots for the ease of exposition. The time slot length $\Delta=\tfrac{D}{N}$ is chosen to be sufficiently small such that the locations of the UAV and the UEs are considered as unchanged within each time slot regardless of the velocity.

\subsection{UE Mobility Model}
We assume that the UEs follow the Gauss-Markov mobility model\cite{liang1999predictive}. Specifically, the velocity of UE $k \in \mathcal{K}$ at time $n+1$ is derived as
\begin{equation}
    \mathbf{v}_k[n+1]=\alpha \mathbf{v}_k[n]+ (1-\alpha)\Bar{\mathbf{v}}+\Bar{\sigma}\sqrt{1-\alpha^2}\mathbf{w}_k[n], 
    \end{equation}
where $\mathbf{v}_k[n]$ is the velocity vector and $\mathbf{w}_k[n]$ is an uncorrelated random Gaussian process $\mathcal{N}(0,\sigma^2)$. Parameters $\alpha, \Bar{\mathbf{v}}$, $\Bar{\sigma}$ represent the memory level, asymptotic mean and asymptotic standard deviation of velocity, respectively \cite{liang1999predictive}. Accordingly, the location is updated as $\mathbf{p}_k[n+1]=\mathbf{p}_k[n]+\mathbf{v}_k[n]\Delta,$ where $\mathbf{p}_k[n] = (x_k[n],y_k[n])$ denotes the location of user $k$ at time $n$, and $\mathbf{p}_k[1]$ denotes its initial location. The UAV is aware of the locations of all UEs at the beginning of each time slot, e.g., from the location feedback by the UEs.

\subsection{Communication Model}
 We assume that the UAV flies at a fixed altitude $h$ during the whole period $D$ with a maximum speed limit $v_m$. Its time-varying horizontal coordinates is denoted as $\mathbf{p}_u[n]=(x_u[n],y_u[n])$ at time slot $n$. The UAV starts from an initial position $\mathbf{p}_u[1] =\mathbf{p}_I$ and is required to reach a predetermined  destination $\mathbf{p}_u[N+1]=\mathbf{p}_F$ at the end of the period. We adopt the commonly used probabilistic line-of-sight (LoS) channel model to determine the large-scale attenuation for UAV-UE links\cite{zeng2019energy,al2014optimal}. The probability of geometrical LoS between the UAV and each UE is dependent on statistical parameters related to the environment and the elevation angle. Specifically, we denote the LoS probability of UE $k$ at time slot $n$ as $\mathbb{P}(LoS,\theta_k[n])$, which can be approximated to be a modified sigmoid function of the following form \cite{al2014optimal} 
 \begin{equation}
 	\mathbb{P}(LoS,\theta_k[n])=\frac{1}{1+a \exp \left(-b(\theta_k[n]-a ) \right)} ,
 \end{equation}
 where $a$ and $b$ are environment-related parameters, and $\theta_k[n]$ is the elevation angle, which is
 \begin{equation}
 	\theta_k[n]= \frac{180}{\pi} \arctan \left(\frac{h}{\norm{\mathbf{p}_u[n]-\mathbf{p}_k[n]} } \right).  
 \end{equation}
 Accordingly, the non-line-of-sight (NLoS) channel probability is equal to $\mathbb{P}(NLoS,\theta_k[n])=1-\mathbb{P}(LoS,\theta_k[n])$. Therefore, the expected channel power gain is 
\begin{align}
    g_k[n]&=\frac{\mathbb{P}(LoS,\theta_k[n])g_0}{d_k[n]^{ \tilde \iota } } +\frac{(1-\mathbb{P}(LoS,\theta_k[n]))\kappa g_0}{d_k[n]^{\tilde\iota} } \notag \\ 
    &=\frac{\hat{\mathbb{P}}( LoS,\theta_k[n])  g_0}{\left(h^2+\norm{\mathbf{p}_u[n]-\mathbf{p}_k[n]}^2\right)^{ \frac{\tilde \iota}{2}}}, \label{eq:gain}
\end{align}
where $\hat{\mathbb{P}}( LoS,\theta_k[n])=\mathbb{P}(LoS,\theta_k[n])+(1-\mathbb{P}(LoS,\theta_k[n]))\kappa $ is the regularized LoS probability by taking into account the attenuation effect of the NLoS channel with $\kappa<1$, $\tilde\iota$ is the path loss exponent, $g_0$ represents the channel gain at the reference distance $d_0=1$ m and $d_k[n]$ is the distance between user $k$ and the UAV at time $n$. During one time slot, we assume that the change of elevation angle $\theta_k[n]$ can be ignored since the UAV movement is relatively small compared with the altitude.  Then, the uplink transmission rate of UE $k$ at time slot $n$ is 
\begin{align}
    R_k[n]&=W\log_2\left(1+\frac{P_kg_k[n]}{N_0}\right)\notag \\&=W\log_2\left(1+\frac{\gamma_k[n]}{\left(h^2+\norm{\mathbf{p}_u[n]-\mathbf{p}_k[n]}^2\right)^{\iota}}\right), \label{eq:rate}
\end{align}
where $\gamma_k[n]= \frac{P_k\hat{\mathbb{P}}( LoS,\theta_k[n])  g_0}{N_0}$, $\iota=\frac{\tilde\iota}{2} $, $P_{k}$ is the fixed transmit power of UE $k$ and $N_0$ is the noise power. Compared to the free-space path loss channel models assumed in the previous work \cite{zhang2018stochastic,zhou2018computation}, the probabilistic model is more general by taking into account the effect of LoS/NLoS channel probability and dealing with more common path loss exponent $\tilde \iota \geq 2$ instead of one special case $\tilde \iota = 2$.
\subsection{Computation Task Model and Execution Methods}
The computation task arrival of each UE is modeled as an i.i.d. Bernoulli process. At the beginning of each time slot, we assume that a computation task with fixed data size $I_k$ (in bits) arrives at user $k \in \mathcal{K}$ with probability $\rho_k$. Denote $A_k[n]$ as the number of arriving bits at time $n$ with $\mathbb{P}(A_k[n]=I_k)=1-\mathbb{P}(A_k[n]=0)=\rho_k$. Each UE maintains a queue for the task arrivals, which will be processed on a FIFO basis. The details of partial computation offloading model are as follows:

    \textit{Local Computing at UE:} The CPU frequency of UE $k$ during time slot $n$ is denoted as $f_k[n]$ (cycles/second). The executed computation bits and the consumed energy within time slot $n$ are given respectively as 
    \begin{align}
        l_k^c[n]&=f_k[n]\Delta/C_k,\\
         E_k^c[n]&=\gamma_c f_k^3[n]\Delta,
    \end{align}
    where $C_k$ is the required number of CPU cycles for computing one bit of input data, and $\gamma_c$ is the effective capacitance coefficient of the processor's chip\cite{bi2018computation}. 
    \par \textit{Computation Offloading:} We assume that the users offload their computation tasks to the UAV using TDMA. The time slot $n$ is further divided into $K$ sub-slots with $\sum_{k=1}^K \delta_k[n]\leq \Delta$, and the $k^{th}$ sub-slot is for UE $k$ to offload its computation task. The offloaded data size and the consumed energy are respectively expressed as
    \begin{align}
        l_k^o[n]&=\delta_k[n] R_k[n],\\
         E_k^o[n]&=\delta_k[n]P_k.
    \end{align}
    
    Therefore, the total number of executed bits and the consumed energy of UE $k$ at time $n$ are given respectively as
    \begin{align}
    	l_k[n]&=l_k^c[n]+l_k^o[n],\\
    	E_k[n]&=E_k^c[n]+E_k^o[n].
    \end{align}

The edge computation time and the feedback downloading time are neglected because the UAV has substantial computation capability and the length of output result is relatively small. As a result, there is no data queue backlog at the UAV. 
    \par  \textit{Task Queue Model:} For $n \in \mathcal{N}$,
  the task queue backlog $Q_k[n]$ evolves as
    \begin{equation}
        Q_k[n+1]=\max\{Q_k[n]+A_k[n]-l_k[n],0\},        \label{eq:qk}
    \end{equation}
    with $Q_k[1]=0, \forall k \in \mathcal{K}$. We refer to the queue $\bm{Q}[n]$ as stable\cite{neely2010stochastic} if:
    \begin{equation}
    	\lim_{N\to \infty} \mathsmaller { \tfrac{1}{N} \sum_{n=1}^N } \mathbb{E}\{\bm{Q}[n]\}<\infty,
     \end{equation}
      where the expectation is taken with respect to the time-varying channels and random task data arrivals.

\subsection{Propulsion Energy Model for Rotary-Wing UAV}
The propulsion energy is the major energy consumption during UAV flight, which is far larger than that consumed on communications and computing. We solicit the existing analytical model for helicopter dynamics, and express the UAV power as a function of velocity 
\begin{equation}
  P_{UAV}(v)=C_1(1+  \tfrac{3v^2}{v_{tip}^2})+C_2\sqrt{\sqrt{C_3+  \tfrac{v^4}{4}} -  \tfrac{v^2}{2}}+C_4v^3,
\end{equation}
where $v_{tip}$ is the tip speed of the rotor, and $C_1, C_2,C_3, C_4$ are constants related to the UAV's weight and its aerodynamic parameters.  During the time slot $n$, the UAV's propulsion energy $E_{UAV}[n]$ is equal to $P_{UAV}[n]\Delta$.

\subsection{Problem Formulation}
Given the upper limit $E^u$ of the UAV's average per-slot energy consumption, we minimize the weighted sum energy consumption of the UEs by jointly optimizing the computation and communication resource allocation and the UAV trajectory. Under random user movements and data arrivals, the problem is formulated as the following multi-stage stochastic optimization problem
\begin{subequations}
\label{eq:p1}
\begin{align}
   \mathcal{P}1: \min_{ \bm{f}[n],\bm{\delta}[n], \atop \mathbf{p}_u[n],    \forall n\in \mathcal{N}}& \lim_{N \to \infty} \mathsmaller{ \tfrac{1}{N}\sum_{n=1}^N\sum_{k=1}^K} w_k E_k[n],\\
   \text{s.t.} \quad& \lim_{N \to \infty} \mathsmaller{\tfrac{1}{N}\sum_{n=1}^N} \mathbb{E}\{ E_{UAV}[n] \} \leq E^u,\\
            \quad& \lim_{N \to \infty} \mathsmaller{\tfrac{1}{N}\sum_{n=1}^N \sum_{k=1}^K} \mathbb{E}\{Q_k[n]\} < \infty,\\
            \quad& 0 \leq f_k[n] \leq f_k^m ,  \forall k,n,\\
            \quad&  \mathsmaller{\sum_{k=1}^K} \delta_k[n] \leq \Delta, \delta_k[n]\geq 0, \forall k,n,\\
            \quad& l_k^c[n]+l_k^o[n] \leq Q_k[n]+A_k[n], \forall k,n,\\
            \quad& \mathbf{p}_u[1]=\mathbf{p}_I, \mathbf{p}_u[N+1]=\mathbf{p}_F,\\
            \quad& ||\mathbf{p}_u[n+1]-\mathbf{p}_u[n]|| \leq v_m\Delta, \forall n,\\
            \quad& 	||\mathbf{p}_F-\mathbf{p}_u[n+1]|| \leq v_m(N-n)\Delta, 
\end{align}
\end{subequations}
where the optimization variables $\bm{f}[n],\bm{\delta}[n],\mathbf{p}_u[n]$ are the combined vector of all UEs' CPU frequencies, the combined vector of all UEs' offloading time and the UAV trajectory at time slot $n$, respectively. (\ref{eq:p1}b) is the long-term UAV propulsion energy constraint. (\ref{eq:p1}c) is the asymptotic queue stability requirement. (\ref{eq:p1}d) and (\ref{eq:p1}e) are constraints on the CPU frequencies and the offloading time. (\ref{eq:p1}f) states that the processed bits cannot exceed the queue backlog plus the data arrival. (\ref{eq:p1}g) - (\ref{eq:p1}i) are the trajectory and speed constraints of the UAV. Without the future knowledge of data arrivals and user locations, we plan to design an online algorithm to solve $\mathcal{P}1$. The non-convex problem $\mathcal{P}1$ has constraints in both spatial and temporal domains. Besides, the trajectory of UAV couples with the offloading computation of the UEs. In the following, we apply the Lyapunov optimization framework to design an online algorithm. 

\section{Lyapunov-based Online Control}
In this section, we apply the Lyapunov optimization to decouple the multi-stage stochastic problem into per-slot deterministic optimization problems that optimize the resource allocation and UAV movement within each time slot.

To cope with the average power consumption constraint in (\ref{eq:p1}b), we introduce a virtual queue as a measurement of the accumulated UAV propulsion energy cost exceeding the required threshold. By setting $Q_u[1]=0$, the virtual queue evolves as
\begin{equation}
\label{eq:qu}
    Q_u[n+1]=\max \left\{Q_u[n]+E_{UAV}[n]-E^u,0 \right\}.
\end{equation}
\par Combining the task queues and the virtual energy queue, we define the \textit{Lyapunov function} as 
\begin{equation}
    L(\bm{Q}[n])=\tfrac{1}{2}(Q_u^2[n]+ \mathsmaller{\sum}_{k=1}^K Q_k^2[n]),
\end{equation}
where $\bm{Q}[n]=\big(Q_u[n],\{Q_k[n]\}_{k=1}^K\big)$ is the concatenated vector of the virtual energy queue and all actual queue backlogs. In practice, we scale the task queues and the virtual queue to be within the similar magnitude to fasten the control process to reach stability. The \textit{conditional Lyapunov drift} is defined as
\begin{equation}
    \Delta L(\bm{Q}[n])=\mathbb{E}\{L(\bm{Q}[n+1])-L(\bm{Q}[n])|\bm{Q}[n]\}.
\end{equation}
Then, the \textit{drift-plus-penalty} \cite{neely2010stochastic} is expressed as
\begin{equation}
    D(\bm{Q}[n])=\Delta L(\bm{Q}[n])+V\mathbb{E}\{E_s[n]|\bm{Q}[n]\},
\end{equation}
where $E_s[n]$ is the weighted sum energy consumption and $V$ is a parameter to control the tradeoff between the system energy cost and the queue stability. 

To minimize the average system energy cost and maintain long-term queue stability, we minimize $D(\bm{Q}[n])$ opportunistically in each time slot $n$. In the following, we first derive an upper bound of $D(\bm{Q}[n])$. 

\textbf{Theorem 1:}  For an arbitrary queue backlog $\bm{Q}[n]$, the drift-plus-penalty is upper bounded as
\begin{align}
       D(\bm{Q}[n]) &\leq \Tilde{B}+Q_u[n]\mathbb{E}
      \{E_{UAV}[n]-E^u|\bm{Q}[n] \}  \notag\\ 
      &+V\mathbb{E}\{E_s[n]|\bm{Q}[n]\}+\mathsmaller{\sum}_{k=1}^K\mathbb{E}\{ Q_k[n]A_k[n] \notag\\ 
      & - (Q_k[n]+A_k[n])l_k[n]|\bm{Q}[n]\}  \label{eq:dpp}
\end{align}
where $\Tilde{B}$ is a finite constant.
\begin{proof}[Proof]
	Squaring the update rules of $Q_k[n]$ \eqref{eq:qk} and $Q_u[n]$ \eqref{eq:qu} and arranging the terms, we have the following inequalities
	\begin{align}
    \frac{Q_u^2[n+1]-Q_u^2[n]}{2}&\leq \frac{1}{2}(E_{\scriptscriptstyle UAV}[n]-E^u)^2 \notag \\
    &+Q_u[n](E_{\scriptscriptstyle UAV}[n]-E^u), \label{eq:eq1} \\
    \frac{ Q_k^2[n+1]-Q_k^2[n]}{2} &\leq \frac{1}{2}(A_k^2[n]+l_k^2[n])+Q_k[n]A_k[n] \notag \\
    &-Q_k[n]l_k[n]-A_k[n]l_k[n], \label{eq:eq2}
    \end{align}
    Taking the conditional expectations of both sides and adding up \eqref{eq:eq1} and \eqref{eq:eq2} over $k$ from 1 to $K$ yields \eqref{eq:dpp},
with $\tilde{B}=\frac{1}{2}\max \{(E^u)^2, (E_{max}-E^u)^2\}+ \sum_{k=1}^K \frac{1}{2}(I_k^2+(f_k^m\Delta/C_k+R_k^m\Delta)^2) $. 
\end{proof}
\par Instead of directly minimizing the drift-plus-penalty, we minimize the upper bound of $D(\bm{Q}[n])$ given in the right-hand-side (RHS) of \eqref{eq:dpp} opportunistically. Specifically, at time slot $n$, we observe the queue state $\bm{Q}[n]$, the arrival tasks $\{A_k[n]\}_{k=1}^K$, the current UAV location $\mathbf{p}_u[n]$, and the user locations $\{ \mathbf{p}_k[n] \}_{k=1}^K$. Accordingly, we control the UAV movement and user task offloading strategies by solving the following optimization problem:
\begin{align}
\label{eq:dop}
    \min_{ \bm{f}[n],\bm{\delta}[n],\mathbf{p}_u[n+1]}\quad &  Q_u[n]E_{UAV}[n] +VE_s[n]\notag\\[-0.5em] 
      & -\mathsmaller{\sum}_{k=1}^K(Q_k[n]+A_k[n])l_k[n], \\
    \text{s.t.} &  \quad\text{(\ref{eq:p1}d)-(\ref{eq:p1}i)}, \notag
\end{align}
where the objective is obtained by eliminating the constant terms in the RHS of \eqref{eq:dpp} given $\bm{Q}[n]$. Thus far, we decouple the original multi-stage optimization problem into a series of deterministic problems in \eqref{eq:dop} for $n \in \mathcal{N}$. For simplicity of illustration, we let $q_k=Q_k[n]+A_k[n]$, drop the time index $n$, and use $\mathbf{p}_{u'}$ to substitute $\mathbf{p}_u[n+1]$. Therefore, the problem is rewritten as follows
\begin{subequations}
\label{eq:p2}
\begin{align}
   \mathcal{P}2:\min_{\bm{f},\bm{\delta},\mathbf{p}_{u'}} & Q_uE_{UAV}-\mathsmaller{\sum}_{k=1}^Kq_k(l_k^c+l_k^o)+VE_s,\\
    \text{s.t.}   \quad& 0 \leq f_k \leq f_k^m,  \forall k,\\
    \quad& \delta_k \geq 0 , \forall k,\\
            \quad& \mathsmaller{\sum}_{k=1}^K \delta_k \leq \Delta, \\
            \quad& l_k^c+l_k^o\leq q_k, \forall k,\\
            \quad& ||\mathbf{p}_{u'}-\mathbf{p}_u|| \leq v_m\Delta, \\
			\quad& 	||\mathbf{p}_F-\mathbf{p}_{u'}|| \leq v_m(N-n)\Delta. 	
\end{align}
\end{subequations}

  \eqref{eq:p2} solves the resource allocation and UAV movement in the current stage, and thus it is an online design without requiring future information.  We summarize our online control strategy in Algorithm 1. 
\begin{algorithm}
 \caption{Lyapunov-based Online Control Algorithm}
    \begin{algorithmic}[1]
    \scriptsize
        \STATE \textbf{Initialization:} $Q_k[1]\leftarrow 0, \forall k, Q_u[1]\leftarrow 0, \mathbf{p}_I,\mathbf{p}_F,v_m$;
        \FOR{$n=1$ to $N$} 
            \STATE Acquire $\bm{Q}[n],\{A_k[n]\}_{k=1}^K,\mathbf{p}_u[n]$, and $\{ \mathbf{p}_k[n] \}_{k=1}^K$;
            \STATE Obtain $\bm{f}^{*}[n],\bm{\delta}^{*}[n]$ and $\mathbf{p}_u^{*}[n+1]$ by solving $\mathcal{P}2$;
            	\FOR{each UE $k$}
            	\STATE Execute $l_k^c[n]$ bits locally using $f_k^{*}[n]$;
            	\STATE Offload data with size $l_k^o[n]$ to UAV during $\delta_k^{*}[n]$;
            \STATE Update user data queue $Q_k[n+1]$ according to \eqref{eq:qk};
            \ENDFOR
            \STATE The UAV provides MEC service to the UEs and flies towards $\mathbf{p}_u^{*}[n+1]$;
            \STATE Update the virtual energy queue $Q_u[n+1]$ according to \eqref{eq:qu}; 
        \ENDFOR
    \end{algorithmic}
\end{algorithm}

\section{Joint Optimization of Resource Allocation and UAV Trajectory}
In this section, we elaborate the joint optimization method to solve the non-convex problem $\mathcal{P}2$. 
 To deal with the non-convexity of the UAV's propulsion energy function, we introduce an auxiliary slack variable $y$ such that
\begin{equation}
    y^2 \geq \sqrt{C_3+\tfrac{v^4}{4}}-\tfrac{v^2}{2}.
    \label{eq:y}
\end{equation}

We introduce another auxiliary variable $\psi_k$ to denote the offloaded bits such that
\[
   \psi_k^2/\delta_k   \leq  W\log_2\left(1+\frac{\gamma_k[n]}{\left(h^2+\norm{\mathbf{p}_u[n]-\mathbf{p}_k[n]}^2\right)^{\iota}}\right).
\label{eq:psi} \addtag
\]

 By substituting \eqref{eq:y} and \eqref{eq:psi} into $\mathcal{P}2$, the problem is transformed to 
\begin{subequations}
\label{eq:rewrite}
\begin{align}
    \min_{\bm{f},\bm{\delta},\mathbf{p}_{u'},\atop y,\psi_k} & 
     Q_u \big(
    C_1(1+\tfrac{3v^2}{v_{tip}^2})+C_2y+C_4v^3 \big) \Delta  \notag \\[-1em]
  & -\mathsmaller{\sum}_{k=1}^K q_k\big(f_k\Delta/C_k+ \psi_k^2 \big) +V E_s,\\
    \text{s.t.}     
            \quad& f_k\Delta/C_k+ \psi_k^2 \leq q_k, \forall k,\\
            \quad&  \psi_k^2/\delta_k \leq R_k, \forall k, \\
            \quad&  C_3/y^2  \leq y^2+ ||\mathbf{p}_{u'}-\mathbf{p}_u||^2/\Delta^2, \\
            \quad& \text{(\ref{eq:p2}b) - (\ref{eq:p2}d),(\ref{eq:p2}f) - (\ref{eq:p2}g)}. \notag
\end{align}
\end{subequations}

The inequality in constraints (\ref{eq:rewrite}c) and (\ref{eq:rewrite}d) must hold at optimum, because otherwise we can decrease the objective without violating the constraint (\ref{eq:rewrite}b) by choosing a smaller $y$ or a larger $\psi_k$. Therefore, \eqref{eq:rewrite} is equivalent to \eqref{eq:p2}. We apply the successive convex approximation (SCA) method to solve problem \eqref{eq:rewrite}. The RHS of (\ref{eq:rewrite}c)  has a concave lower bound as given by Proposition 1.

\noindent  \textbf{Proposition 1:} Given a local point $\mathbf{p}_{u'}^l$ at the $l$-th iteration, the transmission rate of UE $k$ is lower bounded by 
\begin{align}
	R_k^l \{ \mathbf{p}_{u'}\}\triangleq &
     W\log_2\left(1+\frac{\gamma_k}{ \left(h^2+||\mathbf{p}_{u'}^l -\mathbf{p}_k||^2 \right)^{\iota}}\right) \notag \\ &-\beta_k(||\mathbf{p}_{u'}-\mathbf{p}_k||^2-||\mathbf{p}_{u'}^l-\mathbf{p}_k||^2),
     \label{eq:b1}
\end{align}
where $\beta_k=\frac{W(\log_2e)\gamma_k \iota }{[  \gamma_k +(h^2+ ||\mathbf{p}_{u'}^l- \mathbf{p}_k||^2)^{\iota}](h^2+||\mathbf{p}_{u'}^l -\mathbf{p}_k||^2)}$.
\begin{proof}
  Consider the function $f(z)=\log_2(1+\frac{a}{(b+z)^c})$, where $a, b >0, c\geq 1$ and $z \geq 0$. Since $f(z)$ is convex with respect to $z$, its first-order Taylor expansion is a global under-estimator\cite{boyd2004convex}. Given a local point $z_0$, the inequality $f(z) \geq f(z_0)+f^\prime(z_0)(z-z_0)$ holds for any $z$, where $f^\prime(z_0)$ is the derivative the function $f(z)$ at point $z_0$ and $f^\prime(z_0)=\frac{-(\log_2e)ac}{[a+(b+z_0)^c](b+z_0)}$. Therefore, we derive the inequality 
    $\log_2\left(1+\frac{a}{(b+z)^c}\right)\geq \log_2\left(1+\frac{a}{(b+z_0)^c}\right) - \frac{(\log_2e)ac (z-z_0) }{[a+(b+z_0)^c](b+z_0)}.$ With $a=\gamma_k,b=h^2, c=\iota$ and $z_0=||\mathbf{p}_{u'}^l-\mathbf{p}_k||^2$, we obtain the lower bound.
\end{proof}\vspace{-1ex}
 Similarly, we derive a global concave lower bound of the RHS of (\ref{eq:rewrite}d)
\[
 \label{eq:b2}
 Y^l \{ \mathbf{p}_{u'}, y \} \triangleq  (y^l)^2+2y^l(y-y^l)+ \tfrac{||\mathbf{p}_{u'}^l-\mathbf{p}_u ||^2}{\Delta^2} 
    \atop +\tfrac{2}{\Delta^2}(\mathbf{p}_{u'}^l-\mathbf{p}_u)^T(\mathbf{p}_{u'}-\mathbf{p}_u), \addtag
\]
where $y^l$ is defined as 
\[
	y^l=\sqrt{\sqrt{C_3+\tfrac{||\mathbf{p}_{u'}^l-\mathbf{p}_u||^4}{4\Delta^4}}-\tfrac{||\mathbf{p}_{u'}^l-\mathbf{p}_u||^2}{2\Delta^2} }.
\label{eq:yl}
\addtag
\]

 We introduce another auxiliary variable $\theta_k \leq \psi_k^2$, and apply the first-order Taylor expansion. Then, the concave lower bound of $\psi_k^2$ is
\begin{equation}
    \psi_k^2 \geq (\psi_k^l)^2+ 2 \psi_k^l (\psi_k-\psi_k^l) = \Theta^l \{ \psi_k \},
\end{equation}
where $\psi_k^l$ is obtained from \eqref{eq:psi} as
\begin{equation}
\label{eq:psil}
 \psi_k^l= \sqrt{\delta_k^l W\log_2\left(1+\frac{\gamma_k}{ \left(h^2+||\mathbf{p}_{u'}^l -\mathbf{p}_k||^2 \right)^{\iota}}\right) }.
\end{equation}

Consequently, we obtain the convex approximation of $\mathcal{P}2$ in the $l$-th iteration as follows
\begin{subequations}
\begin{align}
   \mathcal{P}3: \min_{\bm{f},\bm{\delta},\mathbf{p}_{u'},\atop y,\psi_k,\theta_k} 
    &Q_u\big(
    C_1(1+\tfrac{3v^2}{v_{tip}^2})+C_2y+C_4v^3 \big) \Delta \notag \\[-1em]
   &-\mathsmaller{\sum}_{k=1}^Kq_k (f_k\Delta/C_k+ \theta_k ) +V E_s,\\
    \text{s.t.}     \quad& f_k\Delta/C_k+ \psi_k^2 \leq q_k, \forall k,\\
            \quad&  \tfrac{\psi_k^2}{\delta_k }  \leq R_k^l \{ \mathbf{p}_{u'} \}, \forall k, \\
            \quad& \tfrac{C_3}{y^2}  \leq Y^l \{ \mathbf{p}_{u'}, y \} , \\
				\quad& \theta_k\leq \Theta^l \{ \psi_k \},\\
            \quad& \text{(\ref{eq:p2}b) - (\ref{eq:p2}d),(\ref{eq:p2}f) - (\ref{eq:p2}g)}. \notag
\end{align}
\end{subequations}

Given the local values, the convex problem $\mathcal{P}3$ can be efficiently solved by off-the-shelf optimization tools such as \texttt{CVX} \cite{boyd2004convex}. The algorithm that jointly optimizes $\mathcal{P}2$ is presented in Algorithm 2.

\begin{algorithm}
\scriptsize
 \caption{SCA-based joint Optimization for $\mathcal{P}2$}
    \begin{algorithmic}[1]
        \STATE \textbf{Input:} A feasible solution \{$\mathbf{p}_{u'}^{(0)}, \bm{\delta}^{(0)}, \bm{f}^{(0)} $\}
			\STATE \textbf{Output:} Resource allocation $\{\bm{f}^{*},\bm{\delta}^{*}\}$; The next UAV location $\mathbf{p}_{u'}^{*}$
 \STATE \textbf{Initialization:} $l\leftarrow 0$, $\epsilon \leftarrow 0.01$;
        \REPEAT
            \STATE Calculate $y^l$ and $\psi_k^l$ according to \eqref{eq:yl} and \eqref{eq:psil}, respectively;

            \STATE Solve the convex problem $\mathcal{P}3$ and denote the optimal values as \{$\bm{f}^{*},\bm{\delta}^{*},\mathbf{p}_{u'}^{*} $\}, denote the objective value as $G^l$; 
            \STATE Update the local values $\mathbf{p}_{u'}^{l+1}=\mathbf{p}_{u'}^{*}, \bm{\delta}^{l+1}=\bm{\delta}^{*}, \bm{f}^{l+1}=\bm{f}^{*} $;
            \STATE Update $l=l+1$;
        \UNTIL $|G^l-G^{l-1}|< \epsilon$
        \STATE \textbf{Return:} $\{\bm{f}^{*},\bm{\delta}^{*},\mathbf{p}_{u'}^{*} \}$
    \end{algorithmic}
\end{algorithm}

\section{Simulation Results}
\begin{figure*}
  \centering  
  \subfigure[UAV Propulsion Energy]{\includegraphics[width=0.32\textwidth]{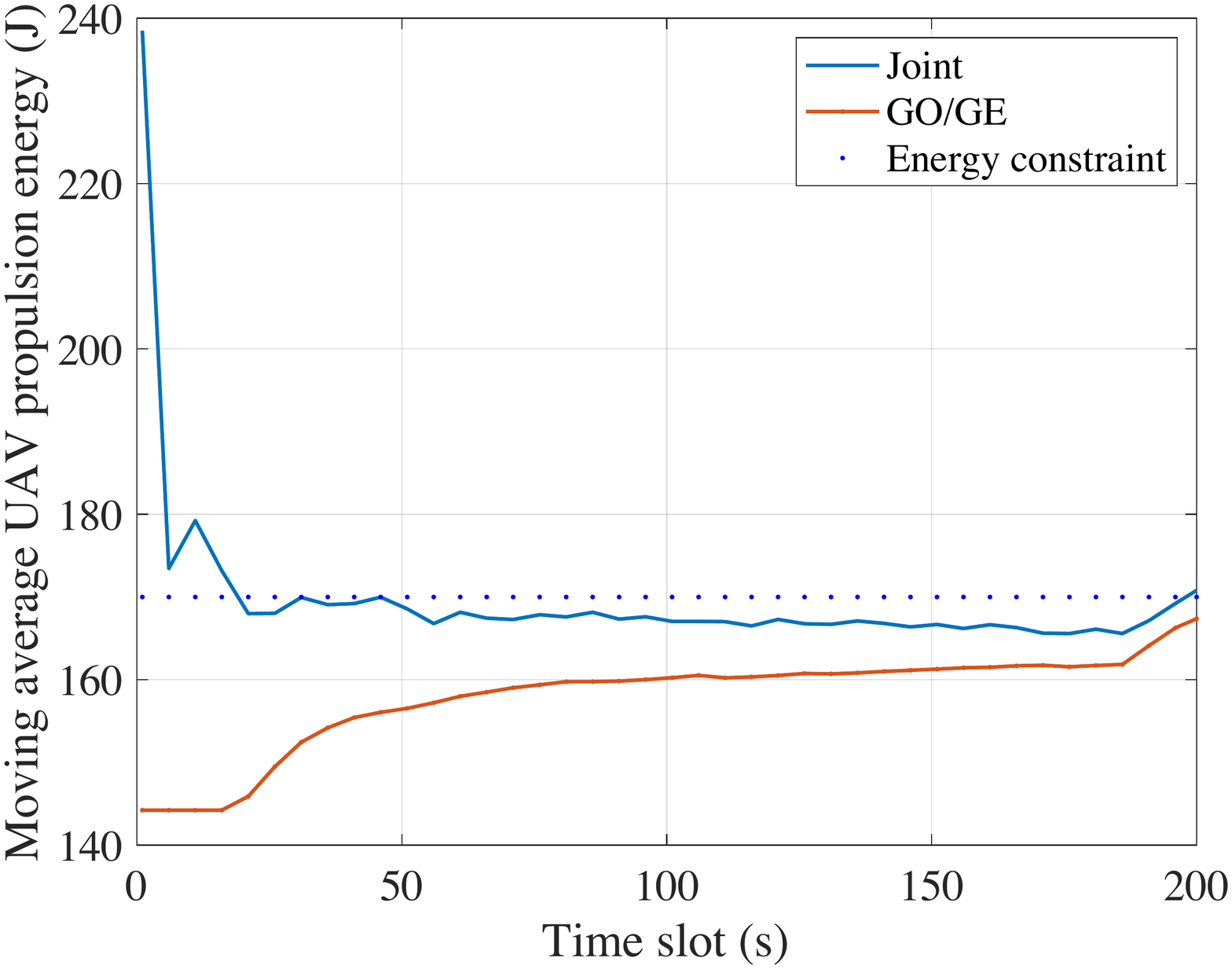}}
  \subfigure[Average user queue length ]{\includegraphics[width=0.32\textwidth]{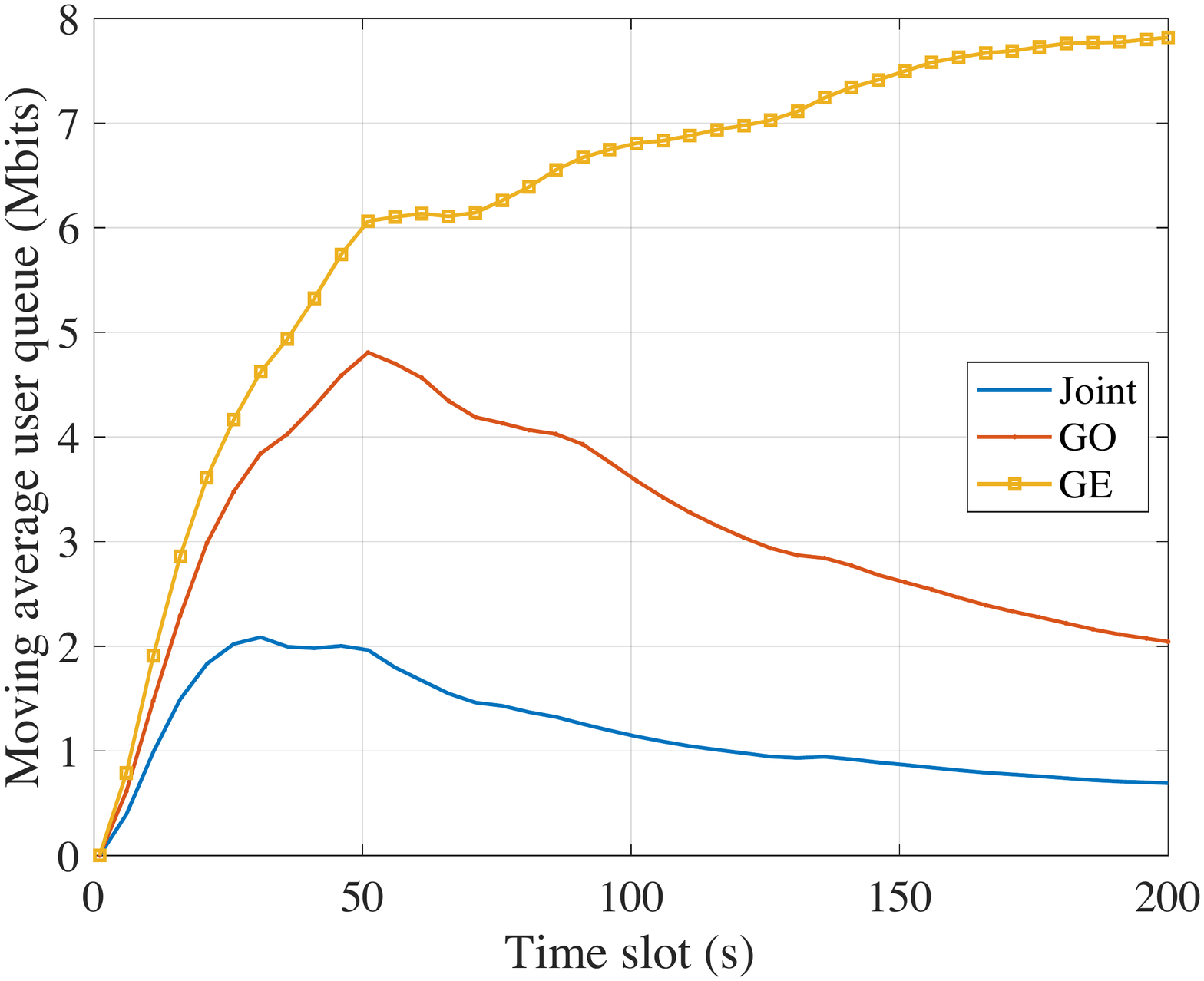}} 
 \subfigure[System energy versus time]{\includegraphics[width=0.32\textwidth]{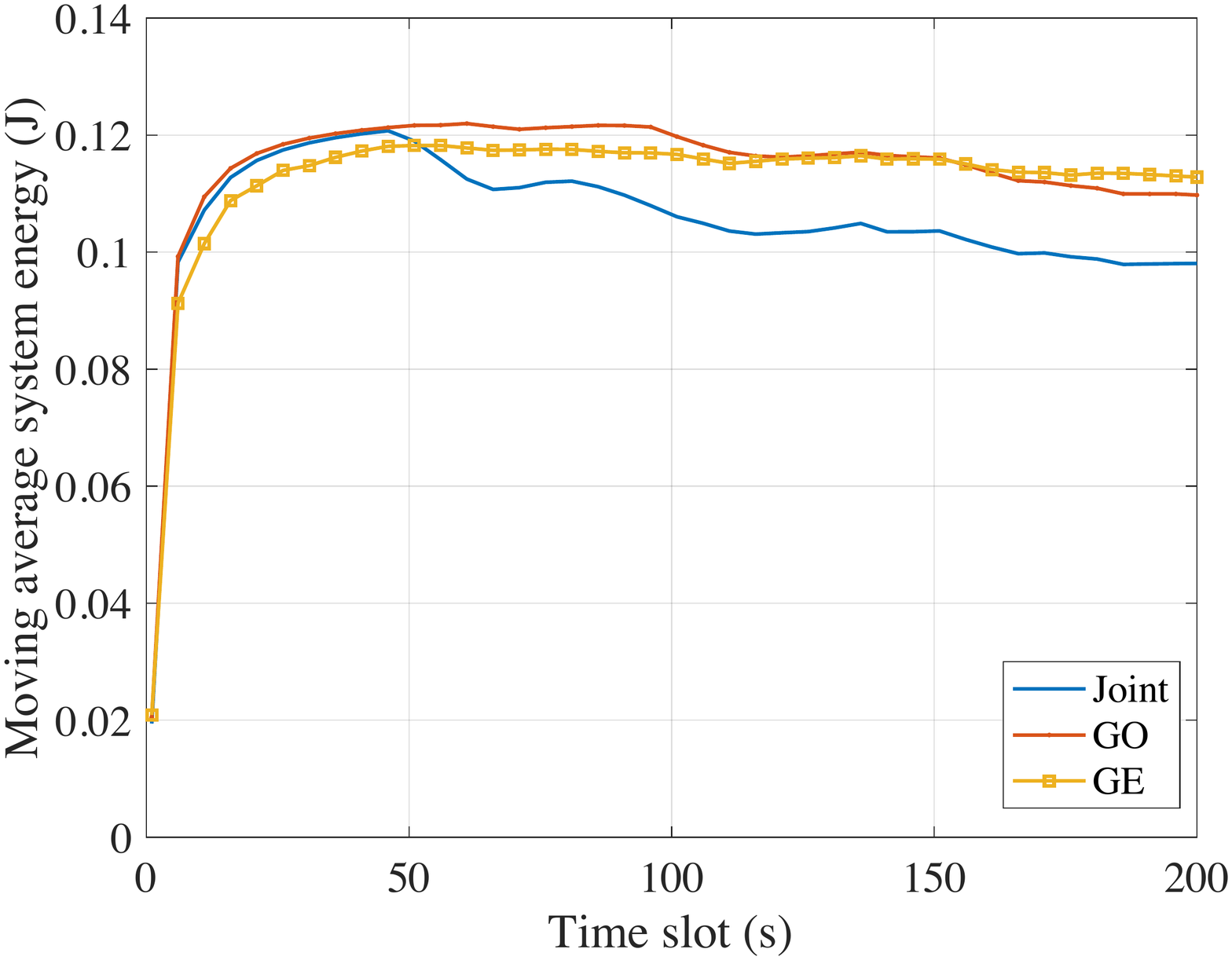}}
  \caption{Convergence performance comparisons of different schemes in the first case.}
  \label{fig:set1}
\end{figure*}

We carry out simulation experiments to evaluate the performance of the proposed methods. As shown in Fig. \ref{fig:set1}(a), we consider the UAV serving four mobile UEs in a 600 m $\times$ 450 m rectangular area. The initial locations of the UEs are 
 $\mathbf{p}_1[1]=[200,100], \mathbf{p}_2[1]=[200,200], \mathbf{p}_3[1]=[200,300]$ and  $\mathbf{p}_4[1]=[200,400]$, respectively. The UEs follow the Gauss-Markov mobility model with $\mathbf{v}[1]= \bar{\mathbf{v}}= [1,0]$ m/s, $\alpha=0.4$, and $\sigma=2$. The initial position and destination of UAV are $\mathbf{p}_I=[0,0]$ and $\mathbf{p}_F=[600,0]$ respectively. It flies at fixed altitude $h=100$ m with maximum velocity $v_m=25$ m/s during $D=200$ s slotted by $N=200$. The related coefficients to calculate the UAV propulsion energy are $C_1=80, C_2=22, C_3=263.4$ and $C_4=0.0092$, respectively. The average UAV energy constraint $E^u$ is equal to $170$ J. The UEs offload their data to the UAV over the channel with bandwidth $W=1$ MHz and noise power $N_0=10^{-12}$ W using fixed transmit power $P_k=0.1$ W. The path loss exponent $\tilde \iota$ is equal to 2.2  and the NLoS attenuation $\kappa$ is 0.2. The reference channel gain $g_0$ is equal to $-50$ dB. The maximum local CPU frequency $f_k^m$ of each UE is 1 GHz. We set the process density $C_k=1000$ cycles/bit and the effective capacitance coefficient $\gamma_c=10^{-28}$. The Lyapunov control parameter $V$ is set to $50$.
 \begin{figure}
	\centering
	\includegraphics[width=0.4\textwidth]{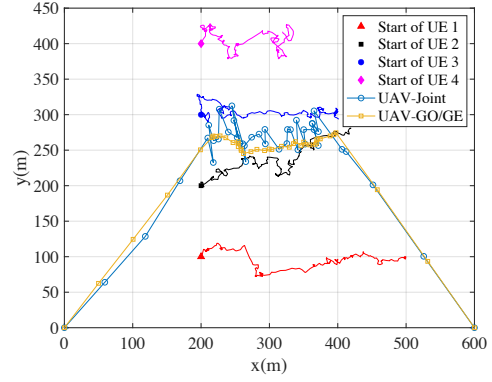}
	\caption{UAV and UEs' trajectories}
	\setlength{\belowcaptionskip}{-10pt}
	\label{fig:tra}
\end{figure}
Besides the proposed joint optimization and the two-stage methods, we also consider two benchmark methods for performance comparison:

 \textit{Geometric Center Tracking + Optimal Resource Allocation (GO):} The UAV keeps tracking the geometric center of all UEs. If the UAV cannot arrive at the center within the current time slot, it will fly towards the center with maximum speed. The computing resource allocation problem is jointly optimized according to the Lyapunov Optimization framework, i.e., following the two-stage method's resource allocation stage.
 
\textit{Geometric Center Tracking + Equal Resource Allocation (GE):} The only difference from the GO method is that the UAV will allocate equal transmission time to UEs that have tasks to process (i.e., non-zero buffer size). Each UE optimizes its local computing frequency and offloading bits according to its assigned period and task queue.

 \par We compare the performance of the three schemes when the task arrivals follow a Bernoulli process with $\mathbb{P}(A_k[n]=2.2)=0.8$ throughout the considered period for all users. In Fig. \ref{fig:tra}, we show the trajectories of four UEs and the projections of UAV under different schemes. The UAV trajectory of the GO and GE methods follow the geometric center of the users. In comparison, the trajectories produced by the joint optimization method vibrates around the UEs' geometric center trajectory due to the random queue backlogs.

 \par We define the moving average energy consumption of the UAV at time slot $n$ as $\tfrac{1}{n}\sum_{\tau=1}^n E_{UAV}[\tau]$. The moving average UE data queue length and the average system energy cost have the similar definitions. We observe from Fig. \ref{fig:set1}(a) that all methods satisfy the average propulsion energy constraint. In Fig. \ref{fig:set1}(b), at the end of the epoch, the average UE queue achieved by the joint method is 0.69M bits, while the queues directed by the GO policy and GE policy are 2.04M bits and 7.82M bits, respectively. Notice from Fig. \ref{fig:set1}(b) that the benchmark policy GO has acceptable performance with a decreasing queue length towards the end of the considered period. In contrast, following the same UAV trajectory as the GO method, the benchmark policy GE cannot stabilize the queue backlog where the queue length increases with time. This demonstrates the importance of Lyapunov control in maintaining data queue stability under the same UAV trajectory policy. In Fig. \ref{fig:set1}(c), the average system energy at the end achieved by the joint method is 0.0981 J. It saves 11.90\% and 15.04\% of the average system energy when compared with the GO policy and GE policy, respectively.

\section{Conclusion}
In this paper, we investigated the long-term average system energy consumption minimization problem in the UAV-enabled MEC system taking dynamic computation offloading, user mobility, resource allocation, and UAV trajectory control into consideration. We adopted the Lyapunov optimization framework to design an online algorithm for the multi-stage stochastic optimization problem. For the non-convex per-slot problem, we proposed a joint optimization method to solve the resource allocation and UAV trajectory control. Simulation results show that the proposed algorithm reduces system energy by tracking the data arrival pattern and UE mobility.

\bibliographystyle{IEEEtran}
\bibliography{IEEEabrv,references}

\begin{thebibliography}{10}
\providecommand{\url}[1]{#1}
\csname url@samestyle\endcsname
\providecommand{\newblock}{\relax}
\providecommand{\bibinfo}[2]{#2}
\providecommand{\BIBentrySTDinterwordspacing}{\spaceskip=0pt\relax}
\providecommand{\BIBentryALTinterwordstretchfactor}{4}
\providecommand{\BIBentryALTinterwordspacing}{\spaceskip=\fontdimen2\font plus
\BIBentryALTinterwordstretchfactor\fontdimen3\font minus
  \fontdimen4\font\relax}
\providecommand{\BIBforeignlanguage}[2]{{%
\expandafter\ifx\csname l@#1\endcsname\relax
\typeout{** WARNING: IEEEtran.bst: No hyphenation pattern has been}%
\typeout{** loaded for the language `#1'. Using the pattern for}%
\typeout{** the default language instead.}%
\else
\language=\csname l@#1\endcsname
\fi
#2}}
\providecommand{\BIBdecl}{\relax}
\BIBdecl

\bibitem{shi2016edge}
W.~Shi, J.~Cao, Q.~Zhang, Y.~Li, and L.~Xu, ``Edge computing: Vision and
  challenges,'' \emph{IEEE internet of things journal}, vol.~3, no.~5, pp.
  637--646, 2016.

\bibitem{bi2018computation}
S.~Bi and Y.~J. Zhang, ``Computation rate maximization for wireless powered
  mobile-edge computing with binary computation offloading,'' \emph{IEEE
  Transactions on Wireless Communications}, vol.~17, no.~6, pp. 4177--4190,
  2018.

\bibitem{mao2017survey}
Y.~Mao, C.~You, J.~Zhang, K.~Huang, and K.~B. Letaief, ``A survey on mobile
  edge computing: The communication perspective,'' \emph{IEEE Communications
  Surveys \& Tutorials}, vol.~19, no.~4, pp. 2322--2358, 2017.

\bibitem{mozaffari2019tutorial}
M.~Mozaffari, W.~Saad, M.~Bennis, Y.-H. Nam, and M.~Debbah, ``A tutorial on
  uavs for wireless networks: Applications, challenges, and open problems,''
  \emph{IEEE Communications Surveys \& Tutorials}, vol.~21, no.~3, pp.
  2334--2360, 2019.

\bibitem{zeng2016wireless}
Y.~Zeng, R.~Zhang, and T.~J. Lim, ``Wireless communications with unmanned
  aerial vehicles: Opportunities and challenges,'' \emph{IEEE Communications
  Magazine}, vol.~54, no.~5, pp. 36--42, 2016.

\bibitem{pham2019survey}
Q.-V. Pham, F.~Fang, V.~N. Ha, M.~Piran, M.~Le, L.~B. Le, W.-J. Hwang, Z.~Ding
  \emph{et~al.}, ``A survey of multi-access edge computing in 5g and beyond:
  Fundamentals, technology integration, and state-of-the-art,'' \emph{arXiv
  preprint arXiv:1906.08452}, 2019.

\bibitem{liu2020path}
Q.~Liu, L.~Shi, L.~Sun, J.~Li, M.~Ding, and F.~Shu, ``Path planning for
  uav-mounted mobile edge computing with deep reinforcement learning,''
  \emph{IEEE Transactions on Vehicular Technology}, 2020.

\bibitem{zhang2018stochastic}
J.~Zhang, L.~Zhou, Q.~Tang, E.~C.-H. Ngai, X.~Hu, H.~Zhao, and J.~Wei,
  ``Stochastic computation offloading and trajectory scheduling for
  uav-assisted mobile edge computing,'' \emph{IEEE Internet of Things Journal},
  vol.~6, no.~2, pp. 3688--3699, 2018.

\bibitem{liang1999predictive}
B.~Liang and Z.~J. Haas, ``Predictive distance-based mobility management for
  pcs networks,'' in \emph{IEEE INFOCOM'99. Conference on Computer
  Communications. Proceedings. Eighteenth Annual Joint Conference of the IEEE
  Computer and Communications Societies. The Future is Now (Cat. No.
  99CH36320)}, vol.~3.\hskip 1em plus 0.5em minus 0.4em\relax IEEE, 1999, pp.
  1377--1384.

\bibitem{zeng2019energy}
Y.~Zeng, J.~Xu, and R.~Zhang, ``Energy minimization for wireless communication
  with rotary-wing uav,'' \emph{IEEE Transactions on Wireless Communications},
  vol.~18, no.~4, pp. 2329--2345, 2019.

\bibitem{al2014optimal}
A.~Al-Hourani, S.~Kandeepan, and S.~Lardner, ``Optimal lap altitude for maximum
  coverage,'' \emph{IEEE Wireless Communications Letters}, vol.~3, no.~6, pp.
  569--572, 2014.

\bibitem{zhou2018computation}
F.~Zhou, Y.~Wu, R.~Q. Hu, and Y.~Qian, ``Computation rate maximization in
  uav-enabled wireless-powered mobile-edge computing systems,'' \emph{IEEE
  Journal on Selected Areas in Communications}, vol.~36, no.~9, pp. 1927--1941,
  2018.

\bibitem{neely2010stochastic}
M.~J. Neely, ``Stochastic network optimization with application to
  communication and queueing systems,'' \emph{Synthesis Lectures on
  Communication Networks}, vol.~3, no.~1, pp. 1--211, 2010.

\bibitem{boyd2004convex}
S.~Boyd and L.~Vandenberghe, \emph{Convex optimization}.\hskip 1em plus 0.5em
  minus 0.4em\relax Cambridge university press, 2004.

\end{thebibliography}

\end{document}